\documentclass[12pt]{iopart}
\usepackage{amssymb}

\usepackage{epsfig}
\usepackage{rotating}
\usepackage{portland}
\usepackage{lscape}
\usepackage{sw20lart}
\usepackage[dvips]{tcigraph}

\input{tcilatex}
\begin{document}

\rotdriver{dvips}

\title{Light deflection in Weyl gravity: constraints on the linear parameter.}
\author{Sophie Pireaux
\footnote{%
Previously working in \newline
Unit\'{e} de Physique Th\'{e}orique et Math\'{e}matique (FYMA),\newline
Universit\'{e} catholique de Louvain (UCL), BELGIUM.}}

\address{UMR 5562, Dynamique Terrestre et Plan\'{e}taire (DTP), B105\newline
Observatoire Midi-Pyr\'{e}n\'{e}es,\newline
14 Avenue Edouard Belin,\newline
31400 Toulouse,\newline
FRANCE}

\ead{sophie.pireaux@cnes.fr}

\begin{abstract}
Light deflection offers an unbiased test of Weyl's gravity since no assumption on 
the conformal factor needs to be made. In this second paper of our series 
``\emph{Light deflection in Weyl gravity}'', we 
analyze the constraints imposed by light deflection experiments on the linear 
parameter of Weyl's theory.
\newline
Regarding solar system experiments, the recent CASSINI Doppler measurements 
are used to infer an upper 
bound, $\sim 10^{-19}\ $m$^{-1}$, on the absolute value of the above Weyl 
parameter. In non-solar system experiments, a condition for unbound orbits 
together with gravitational mirage observations enable us to further 
constrain the allowed negative range of the Weyl parameter to 
$\sim -10^{-31}\ $m$^{-1}$.
\newline
We show that the characteristics of the light curve in microlensing or gravitational 
mirages, deduced from the lens equation, cannot be recast into the General Relativistic 
predictions by a simple rescaling of the deflector mass or of the ring radius. 
However, the corrective factor, which depends on the Weyl 
parameter value and on the lensing configuration, is small, even perhaps negligible, 
owing to the upper bound inferred on the absolute value of a negative Weyl parameter. 
A statistical study on observed lensing systems is required to settle the question.

Our Weyl parameter range is more reliable than the single value derived by Mannheim and 
Kazanas from fits to galactic rotation curves, $\sim +10^{-26}\ $m$^{-1}$. 
Indeed, the latter, although consistent with our bounds, is biased by the choice of a specific
conformal factor.
\end{abstract}

\submitto{\CQG}

\pagebreak

\section{Introduction}

\qquad As explained in reference $\cite{Pireaux 2003 critical
distances in Weyl}$ an alternative theory of gravitation is highly
desirable. Indeed, solely from a theoretical point of view, the choice of
the Hilbert-Einstein action is not based on any fundamental principle,
Einstein's theory of gravitation cannot be properly described by quantum
field perturbation theory, neither is it invariant under conformal
transformations. Those latter two lacks make it difficult to unify
gravitation with other fundamental interactions. From the point of view of
experiments, the Newtonian potential recovered in the weak field regime of General
Relativity cannot reproduce the flat velocity distributions in the vicinity
of galaxies without copious amounts of dark matter.

Regarding the demand for an alternative theory, and among many
candidates, the Weyl theory is an interesting prototype. Not only is it
conformally invariant, but it contains an additional linear
contribution to the Newtonian potential. This latter feature is encoded in
the key parameter of the theory, $\gamma _{W}$. The Weyl gravity holds a total 
of three free parameters: $\gamma _{W}$, $\beta _{W}$ and $k_{W}$; plus a 
conformal factor which is to be specified by a consistent study of the coupling 
of the Weyl gravity to matter fields. To recover a Newtonian potential for
photons on short distance scales, the second parameter is constrained to $%
\beta _{W}(M)=\frac{G_{N}M}{c^{2}}$, where $G_{N}$ is the Newtonian
constant; $M$, the total gravitational mass (luminous or not); and $c$, 
the speed of light. General Relativity then corresponds to the particular case
$\gamma _{W}=k _{W}=0$. The parameter $k _{W}$ should be effective only 
on cosmological distance scales.
Nevertheless, as shown in reference $\cite{Pireaux 2003 critical
distances in Weyl}$, photon paths are insensitive to $k_{W}$, as well as
to the unknown conformal factor. Hence, light deflection experiments offer
an interesting tool to constrain the Weyl linear parameter $\gamma _{W}$. In
the following approach, we shall call it the Weyl parameter and neglected
the mixed ($\beta _{W}\gamma _{W}$)-term when compared with that of $\gamma
_{W}$ and $\beta _{W}$ alone.

In reference $\cite{Pireaux 2003 critical distances in Weyl}$,
we inferred criteria for light deflection to take place and introduced 
critical distances in Weyl gravity regarding photon trajectories. Let us
recall the relevant ones, namely: the distance that separates between
unbound and bound orbits, 
\begin{equation}
r_{null}
\begin{array}[t]{cl}
\sim  & -\frac{1}{\gamma _{W}}
\end{array}
\text{ ,}  \label{r_null_Weyl}
\end{equation}
which is physical for negative values of the Weyl parameter; and the
critical closest approach distance separating the convergent and the divergent regime, 
\begin{equation}
r_{0\ 0}
\begin{array}[t]{cl}
\sim  & \sqrt{6} \cdot 10^{\frac{x+3}{2}}\frac{1}{\sqrt{\left| \gamma _{W}\right| 
}}\ \text{m for }M=10^{x}M_{Sun}\text{, }\gamma _{W}\text{ in }\left[ \text{m%
}^{-1}\right] \text{ ,}
\end{array}
\label{r0_0_Weyl}
\end{equation}
when the Weyl parameter is positive. As in article $\cite{Pireaux 2003
critical distances in Weyl}$, our present analysis will be carried out
consistently with respect to the weak field 
\begin{equation}
r_{weak\ field}
\begin{array}[t]{cl}
\ll  & \frac{1}{\left| \gamma _{W}\right|}
\end{array}
\label{weak_field_radius_Weyl}
\end{equation}
and the strong field 
\begin{equation}
r_{strong\ field}
\begin{array}[t]{cl}
\ggg  & \sqrt{3} \cdot 10^{\frac{x+3}{2}}\frac{1}{\sqrt{\left| \gamma _{W}\right| 
}}\ \text{m for }M=10^{x}M_{Sun}\text{, }\gamma _{W}\text{ in }\left[ \text{m%
}^{-1}\right] \medskip 
\end{array}
\label{strong_field_radius_Weyl}
\end{equation}
approximations previously introduced. Indeed, together with the critical
radii for photons, those limiting distances are functions of the linear 
parameter $\gamma _{W}$.

Keeping those key distances in mind, we shall now confront predictions
of Weyl theory regarding light deflection with observations in order to
constrain $\gamma _{W}$. Historically, the spectacular phenomenon of light
deflection by a gravitational source (namely the Sun) was the trigger to the
success of Einstein's theory. Today, the arrival of new detection techniques
allows not only for precise measurement of the light deflection angle
(change in the apparent position of a light source) due to the Sun or
planets in our Solar System, but also for observation of microlensing
events (variation of the received light flux) at the scale of our galaxy,
and of gravitational mirages (multiple images) or weak lensing (distorted
images) at extragalactic distance scales. So light deflection experiments
allow us to test the universality of relativistic theories of gravitation
over different distance scales by considering successively a close star, a
galaxy or a distant quasar as light sources lensed by a
Solar System body, a star of the galactic halo, a galaxy or a cluster. Doing
so, they also explore very different mass scales for the lens.

\section{Solar System experiments}

\qquad The present section of this article aims at a confrontation of Weyl
gravity with Solar System experiments. In the previous article ``\emph{Light
deflection in Weyl gravity: critical distances for photon paths}'' by S.
Pireaux $\cite{Pireaux 2003 critical distances in Weyl}$, we obtained
the expression for the asymptotic light deflection angle in the weak field
regime (Equation (18) in $\cite{Pireaux 2003 critical distances in Weyl}$). %
Neglecting the ($\beta _{W}\ \gamma _{W}$)-contribution in the Weyl
gravitational potential (Equation (4) in $\cite{Pireaux 2003 critical
distances in Weyl}$), we find
\begin{equation}
\hat{\alpha}_{\text{weak field}}(r_{0})\simeq +\frac{4\ \beta _{W}\ }{r_{0}}%
-\gamma _{W}\ r_{0}  \label{angle_deflex_weak_above_parsec_WEYL}
\end{equation}
where $r_{0}$ is the closest approach distance of the photon to the
deflector.
The prediction about the weak field deflection angle already
differs from the Einsteinian one as soon as $\gamma _{W}$ is nonzero. 

The light deflection angle due to the gravitational field of the Sun was the
very first prediction of General Relativity which originally confirmed this
theory within a 20\% error margin. Today, with modern techniques operating
in the radio-waveband, the precision has reached about 0.001\% 
\cite{Bertotti 2003 Cassini gamma measurement}, allowing the
planets of our Solar System to be considered as potential deflectors too. At
first order, deviations from General Relativity regarding light deflection
are encoded in the Post-Newtonian (PN) parameter $\gamma $ (not to be confused
with the Weyl parameter $\gamma _{W}$), with $\gamma =1$ for General
Relativity.

\subsection{Estimation of $\gamma _{W}$ from VLBI data}

\qquad We use first order light deflection measurements provided by
Very Large Baseline Interferometry (VLBI), more precisely measurements of
the corresponding time delay due to photons being deflected by the Sun, to
obtain constraints on the Weyl parameter $\gamma _{W}$.\newline
The measurements dedicated to $\gamma $ made in 1995 using quasars 3C273 and 
3C279 $\cite{Lebach 1995 VLBI}$ constrained the Post-Newtonian parameter $\gamma $ to 
\begin{equation}
\gamma =0.9996\pm 0.0017\,.  \label{estimated_gamma_VLBI_1}
\end{equation}
Now, VLBI type II experiments are in progress, no more dedicated to light
deflection, but providing an indirect measurement of $\gamma $, while
monitoring polar motion and Earth rotation. The project has grown into a
network of more than 87 observatories and is sensitive to light deflection
over almost the entire celestial sphere. An analysis of over 1.7 million
ionosphere-corrected group delay measurements involving 541 radio sources 
$\cite{Shapiro 2004 VLBI estimate of gamma}$ lead to the value of\footnote{%
An analysis of over 2 million observations quoted by Will 
$\cite{Will 2001 summary of tests}$ from
reference $\cite{Eubanks 1999 advances in solar system tests}$ lead to 
$(1+\gamma )/2=0.99992\ \pm 0.00014\,.$
But reference $\cite{Eubanks 1999 advances in solar system tests}$ 
is unpublished. Other less stringent estimates of $\gamma $ where obtained 
in the past from VLBI data, see $\cite{Seielstad 1970 VLBI gamma determination}$, 
$\cite{Counselman 1974 VLBI gamma determination}$, 
$\cite{Fomalont 1976 VLBI gamma determination}$, 
$\cite{Robertson 1984 VLBI gamma measurements}$, 
$\cite{Robertson 1991 VLBI gamma measurements}$ and  
$\cite{Robertson 1991 VLBI gamma measurements proceedings}$.}  
\begin{equation}
\gamma =0.99983\ \pm 0.00045\,.  \label{estimated_gamma_VLBI_2}
\end{equation}

We now assume $\beta _{W}$ given by $\frac{G_{N}M_{Sun}%
}{c^{2}}$ to recover the Newtonian potential in the Solar System
weak field limit, and extrapolate the VLBI results for light
deflection at the solar limb. Matching the first order Post-Newtonian
expression for the light deflection angle with that of the Weyl theory (\ref
{angle_deflex_weak_above_parsec_WEYL}) leads to\footnote{%
Even though, strictly speaking, the Weyl gravity does not allow for a
Post(-Post) Newtonian development, because the corresponding potential for
photons diverges on asymptotical radial distances.} 
\begin{equation}
\left. 
\begin{array}[b]{c}
\gamma _{W}=\frac{\left( 1-\gamma \right) }{2}\frac{4G_{N}M_{Sun}}{%
R_{Sun}c^{2}}\frac{1}{R_{Sun}} \\ 
\Downarrow  \\ 
\left\{ 
\begin{array}{ll}
-7.9 \cdot 10^{-18}\ \text{m}^{-1}\leq \gamma _{W}\leq 1.3 \cdot 10^{-17}\ \text{m}%
^{-1} & \text{for (\ref{estimated_gamma_VLBI_1}),} \\ 
-1.7 \cdot 10^{-18}\ \text{m}^{-1}\leq \gamma _{W}\leq 3.8 \cdot 10^{-18}\ \text{m}%
^{-1} & \text{for (\ref{estimated_gamma_VLBI_2}).}
\end{array}
\right. 
\end{array}
\right.   \label{estimated_value_of_gamma_Weyl}
\end{equation}
This estimation with recent VLBI data provides a range of values for $\gamma
_{W}$ which contains the particular order of magnitude needed by Mannheim
and his collaborators to fit the galactic rotation curves $\cite{Mannheim
1994 microlensing}$, $\cite{Mannheim 1995 Age of Universe}$, $\cite{Mannheim
1989 Exact Solution}$, namely
\begin{equation}
\gamma _{W\text{ Mannheim-Kazanas}}\sim +10^{-26}\ \text{m}^{-1}\,.
\label{parametrization_Mannheim_Kazanas}
\end{equation}
The range (\ref{estimated_value_of_gamma_Weyl}) is narrower than that given by 
Edery \emph{et al}. $ \cite{Edery 1998 lensing in Weyl theory}$, based on an 
estimation of the PN parameter given in an article of 1976.

\subsection{Estimation of $\gamma _{W}$ from Cassini data}

\qquad The Cassini experiment was carried out between 6th of June and 7th of
July 2002. The spacecraft was on its way to Saturn. Measurements were
made around the time of solar conjunction, at which the spacecraft was almost
aligned with the Sun and the Earth, that occured on 21st June 2002. Those
new constraints on the PN parameter $\gamma $ were obtained with a Doppler
method. Motion of the spacecraft produces a change in the time delay of
light transmited between the spacecraft and the Earth, as well as in its
impact parameter respective to the Sun. Those are equivalent to a change in
the distance and hence in the relative radial velocity between the
spacecraft and the Earth, resulting in a Doppler effect. The Doppler method
coupled to a new radio configuration using double-band and multifrequency
link allowed to increase the constraints on $\gamma $ by one order of
magnitude \cite{Bertotti 2003 Cassini gamma measurement}. 
The constraints we infer on the Weyl parameter are correspondingly
improved:

\begin{equation}
\begin{array}{ll}
\gamma -1=(-2.1\pm 2.3)\cdot 10^{-5} & 
\end{array}
\label{estimated_gamma_Cassini_2}
\end{equation}

\begin{equation}
\left. 
\begin{array}{c}
\Downarrow \\ 
\left. 
\begin{array}{ll}
-1.2 \cdot 10^{-20}\ \text{m}^{-1}\leq \gamma _{W}\leq 2.7 \cdot 10^{-19}\ \text{m}%
^{-1} & \text{for (\ref{estimated_gamma_Cassini_2}).}
\end{array}
\right.
\end{array}
\right.  \label{estimated_value_of_gamma_Weyl_Cassini}
\end{equation}

\subsection{Sign of $\gamma _{W}$?}

\qquad There exist planed experiments which should improve measurements
of the Post-Newtonian parameter $\gamma $ with a precision of $\sim 5
\cdot 10^{-7}$, like the future GAIA mission intended to measure $\gamma $ as a
by-product of microarcsecond-astrometry $\cite{GAIA 2000 study repport}$; or
the project LATOR, dedicated to $\gamma $, which should reach a precision of 
$\sim 5 \cdot 10^{-8}$ $\cite{Turyshev 2003 LATOR}$. Although higher precision
tests on light deflection could reduce the allowed range of values for $%
\gamma _{W}$, the test of light deflection in the neighborhood of the Sun
cannot help us to decide on the sign of the parameter $\gamma _{W}$.

Considering that the light deflection angle calculated with 
General Relativity for the visible gravitational mass in galaxies or
clusters at galactic distance scales is often inferior to the observed
deflection, the presence of gravitational dark matter is inferred. 
So, if one wishes the
linear $\gamma _{W}$-term of the Weyl potential to be an alternative to (a
too large amount of) dark matter contributing to light deflection, then $%
-\gamma _{W}\ r_{0}$ must be positive. 
However, this sign of $\gamma _{W}$ is just the opposite of the sign argued by
Mannheim and Kazanas in their parametrization (\ref
{parametrization_Mannheim_Kazanas}). Two types of arguments might be given
in order to solve this apparent contradiction and prevent ruling out
the Weyl theory.\newline
A first possibility is to consider $\gamma _{W}$ to be positive, so that the
theory would still need the ``magic'' contribution of dark matter to explain
light deflection due to galaxies and clusters, just like General Relativity
does. This possibility leads to divergent deflection (a negative $\widehat{%
\alpha }$ in Equation (\ref{angle_deflex_weak_above_parsec_WEYL})) on radial
closest approach distances from the deflector larger than $r_{0\ 0}(M,\gamma
_{W})$ (\ref{r0_0_Weyl}), where the linear divergent contribution becomes
dominant. The positive sign was considered in our graduate thesis work $\cite
{Pireaux 1997 memoire}$, without dark matter.\newline
An alternative argument would be to claim that tests of the Weyl theory
taking into account massive particles or bodies are ambiguous, in comparison
with those based on nonmassive particles like photons, or ultra relativistic
particles. Indeed, the presence of matter breaks the conformal symmetry of
the theory, and this symmetry breaking mechanism is not well understood. In
order to fit galactic rotation curves, we need to fix the arbitrary
conformal factor of the line element (Equation (3) in $\cite{Pireaux 2003
critical distances in Weyl}$) because massive geodesics are not conformally
invariant. Mannheim and Kazanas (arbitrarily) chose the conformal factor $%
\chi ^{2}(r)\equiv 1$ (or constant) and obtained $\gamma _{W}>0$ through
those fits. Their work corresponds to a particular theory:

\begin{equation}
\fl\fbox{{\small Mannheim-Kazanas theory}$\,\equiv \,\left\{ 
\begin{array}{l}
\text{{\small Weyl theory}} \\ 
+ \\ 
\chi ^{2}(r)\equiv 1 \\ 
+ \\ 
\text{{\small fits to galactic rotation curves} {\small (\ref
{parametrization_Mannheim_Kazanas})}}
\end{array}
\right\} $}  \label{definition_Mannheim_theory}
\end{equation}

However, the physical conformal factor in the spherically symmetric metric
(Equation (3) in $\cite{Pireaux 2003 critical distances in Weyl}$),
specified by the symmetry breaking mechanism, could be different from a
constant. Hence, the physical parameter $\gamma _{W}$ present in the metric
would be different from the estimate of Mannheim and Kazanas. For example,
Edery \emph{et al}. $\cite{Edery 2000 weyl gauge choice}$ have shown that,
in the weak field limit (which applies to the galactic rotation-curve
parametrization and to light deflection in the Solar System), it is possible
to find an appropriate conformal factor $\chi ^{2}(r)$ and a radial
coordinate transformation $r^{^{\prime }}(r)$, so as to change the sign of
the $\gamma _{W}$-term in the Weyl gravitational potential $V_{W}(r,\gamma
_{W})$ given in $\cite{Pireaux 2003 critical distances in Weyl}$, when the ($%
\beta _{W}\ \gamma _{W}$)-term is neglected. One easily checks
that, if the conformal factor is given by $\chi
^{2}(r)\simeq 1-2\gamma _{W}\ r$ with $r^{^{\prime }}=\chi (r)\cdot r$, then
\begin{eqnarray}
ds^{2}(r,\gamma _{W}) &\equiv &\chi ^{2}(r)\cdot \left\{ 
\begin{array}{l}
+\left[ 1+\frac{2\,V_{W}(r,\gamma _{W})}{c^{2}}\right] c^{2}dt^{2} \\ 
-\left[ 1+\frac{2\,V_{W}(r,\gamma _{W})}{c^{2}}\right] ^{-1}dr^{2} \\ 
-r^{2}\left( d\theta ^{2}+\sin ^{2}\theta \,d\varphi ^{2}\right) 
\end{array}
\right\} \smallskip   \nonumber \\
&\simeq &\left\{ 
\begin{array}{l}
+\left[ 1+\frac{2\,V_{W}(r^{^{\prime }},\gamma _{W}^{^{\prime }})}{c^{2}}%
\right] c^{2}dt^{2} \\ 
-\left[ 1+\frac{2\,V_{W}(r^{^{\prime }},\gamma _{W}^{^{\prime }})}{c^{2}}%
\right] ^{-1}dr^{^{\prime 2}} \\ 
-r^{^{\prime }2}\left( d\theta ^{2}+\sin ^{2}\theta \,d\varphi ^{2}\right) 
\end{array}
\right\} \equiv ds^{2}(r^{^{\prime }},\gamma _{W}^{^{\prime }})  \nonumber \\
&&\text{\bigskip }  \nonumber \\
&&\text{where }\gamma _{W}^{^{\prime }}\equiv -\gamma _{W}\,.
\label{conform_transfo_change_sign_Weyl}
\end{eqnarray}

In conclusion, until a conformal factor is specified by a consistent study of
the coupling of Weyl gravity to matter fields, the conservative bounds
deduced from Solar System light deflection experiments (\ref
{estimated_value_of_gamma_Weyl}, \ref{estimated_value_of_gamma_Weyl_Cassini}%
) are preferable.\newline
And if a negative sign is taken for $\gamma _{W}$ for photons, then light
deflection is always convergent.

\section{Beyond Solar System experiments}

\subsection{Constraints on a negative $\gamma _{W}$ from the existence of
gravitational mirages}

\qquad We have shown in the article ``\emph{Light deflection in Weyl
gravity: critical distances for photon paths}'', in particular through
discussion (22) and Figure 9 of reference $\cite{Pireaux 2003
critical distances in Weyl}$, that there exists no asymptotic state for
photons if the closest approach distance $r_{0}$ is larger than $r_{null}$ (%
\ref{r_null_Weyl}), when the Weyl linear parameter $\mathbf{\gamma }_{%
\mathbf{W}}$\textbf{\ is negative}. Hence no asymptotic light deflection can take
place. The existence of gravitational mirages with an Einstein angle $%
\vartheta _{E}$ (corresponding to the Einstein radius $r_{E}$) of a few
arcseconds puts a stronger upper bound on the absolute value of the Weyl
linear parameter. Indeed, using a rough estimation of the distance as a function 
of the redshift, based on the empirical Hubble Law, 
\begin{equation}
D\simeq \frac{c}{H_{0}}\ z\quad \text{for }z<1\,,
\label{rough_estimate_redshift_from_distance}
\end{equation}
we find 
\begin{equation}
\begin{array}{ll}
\frac{r_{0}}{Dol}\sim \frac{r_{E}}{Dol}\simeq \vartheta _{E}\lesssim \frac{%
r_{null}}{Dol}\bigskip \\ 
\stackrel{\text{{\tiny (\ref{r_null_Weyl})}}}{\Leftrightarrow } \left|
\gamma _{W}\right| \lesssim \left[ \frac{1}{\vartheta _{E}}\ h_{0}\ \frac{%
0.3}{z_{L}}\right] \ 1.7 \cdot 10^{-31}\ \text{m}^{-1}\quad \text{for }\gamma
_{W}<0 \\ 
& \smallskip \\ 
\fl \text{with }\left\{ 
\begin{array}{l}
D_{ol},\text{ the distance observer-lens,} \\ 
\vartheta _{E},\text{ the Einstein angle corresponding to a mirage ring, in }%
\left[ \text{arcsec}\right] \text{,} \\ 
h_{0},\text{ the normalized Hubble parameter }\in \left] 0,55;0,75\right[ \\ 
\quad \text{with }H_{0}=100\ h_{0}\ \text{km}\ \text{s}^{-1}\ \text{Mpc}^{-1}%
\text{ }\cite{Turner 1999 Estimations of Omega}\text{,} \\ 
z_{L},\text{ the redshift of the lens,}
\end{array}
\right. \smallskip
\end{array}
\label{statistic_estimation_of_gamma_negative_mirages_Weyl}
\end{equation}
Consequently, it seems reasonable to admit the following conservative limit
on $\gamma _{W}:$%
\begin{equation}
\left| \gamma _{W}\right| <\ 10^{-31}\ \text{m}^{-1}\text{\quad for }\gamma
_{W}<0\,\text{.}  \label{estimated_value_of_gamma_negative_Weyl}
\end{equation}

When $\mathbf{\gamma }_{\mathbf{W}}$\textbf{\ is positive}, on the contrary,
we cannot find a better upperbound than the conservative value obtained from Solar System
experiments (\ref{estimated_value_of_gamma_Weyl_Cassini}).  
Otherwise, we have to adopt the Mannheim-Kazanas parametrization (\ref
{parametrization_Mannheim_Kazanas}) based on an arbitrary constant value of
the conformal factor.

\subsection{Relevance of the weak field versus the strong field limit}

\qquad \label{discussion_pertinence_weakfield_limit_W}If we wish to infer 
further constraints on the Weyl parameter from microlensing or
gravitational lensing events, it is crucial to stick to the constraints so
far available, respectively (\ref
{estimated_value_of_gamma_negative_Weyl}) for a negative parameter, or (\ref
{estimated_value_of_gamma_Weyl_Cassini}) for a positive parameter, and to
respect the limits of the approximations introduced in reference $%
\cite{Pireaux 2003 critical distances in Weyl}$ which are functions of $\gamma
_{W}$.
In fact, the weak (or strong) field limit on the radial
distance measured from the gravitational lens has to be verified on the
photon path, all the way from the light source to the observer. That is, 
the limit does not only apply to the lens-observer ($Dol$) and 
lens-source ($Dls$) distances, but also to the closest approach distance 
of the photon onto the lens ($r_{0} $). 
This distance is considered to be of the same order of magnitude 
as the Einstein ring radius ($r_{E}$) associated with the
Observer-Lens-Source (O-L-S) system.\medskip

Immediately, we see that \textbf{the strong field limit} (\ref
{strong_field_radius_Weyl}) is of no use here, because we only
have in hand an upper bound on $\left| \gamma _{W}\right| $. The lower bound
is given by General relativity ($\gamma _{W}=0 $) and leads to a strong
field limit only valid at infinity.\medskip

As far as the \textbf{weak field limit} (\ref{weak_field_radius_Weyl}) is
concerned, we are limited by our upper bound on $\left| \gamma
_{W}\right| $.

Our numerical example for a \underline{microlens} is realized for a
point-like lens (L) of one solar mass placed in the galactic cloud, and a
stellar source (S) present in the Large Magellanic Cloud: 
when looking towards the halo of our Galaxy, an O-L-S microlensing
system characterized by
\begin{eqnarray}
Dos &=&2 \cdot 10^{21}\ \text{m}\,,  \nonumber \\
Dol &=&5 \cdot 10^{20}\ \text{m}\,\text{,}  \label{simulation_num_microlens}
\end{eqnarray}
which imply 
\begin{equation}
r_{E}=1.5 \cdot 10^{12}\ \text{m ,}
\label{radius_Einstein_ring_microlensing_simulation}
\end{equation}
and a typical angular separation between images of the order of the
milliarcsecond. Alternatively, if one was considering microlensing towards
the Galactic Bulge, the characteristic distances would be instead $Dos\sim
8\ $kpc$=2.5 \cdot 10^{20}\ $m and $Dol=Dos/2$. For a lens of one solar mass,
this means an Einstein radius of $6.1 \cdot 10^{11}\ $m.
Microlensing by a point mass model is a good approximation of reality in 
the simple limiting case of small lensing probability along the
line of sight. A microlensing event in the massive halo of the
Milky Way is a good example of this. On the contrary, if the line of sight
passes through the center of the galaxy, the lensing optical depth (lensing
probability) may approach $1$, requiring a more complex model with
intricate mass contributions.
In the case of microlenses in the dark halo, the lens speed is negligible
with respect to the speed of light. Therefore, we ignore the frequency
shift of the light due to the changing of path length along the 
lines of sight for the different images. Hence, the surface brightness is
the same for all the images, and the flux density is proportional to the
solid angle of the images. This point allows easy computation of the
image amplification.
\newline
Owing to distance scales involved in
microlensing events, the conservative estimate (\ref
{estimated_value_of_gamma_Weyl_Cassini}) from Solar System
experiments, will not allow us to improve constraints on a positive
parameter, because, strictly speaking, we are not allowed to use this weak
field limit. On the contrary, the more stringent bound (\ref
{estimated_value_of_gamma_negative_Weyl}) obtained for a
negative parameter makes it possible to discuss microlensing predictions
when $\gamma _{W}$ is negative.

In the case of \underline{gravitational mirages}, we shall use a point-like
lens model representing either a galaxy ($M\sim 10^{11}\ M_{Sun}$) or a
cluster of galaxies ($M\sim 10^{13}-10^{14}-10^{15}\ M_{Sun}$), with the
following distance scales
\begin{eqnarray}
Dol &=&10^{8}\frac{G_{N}M}{c^{2}}\,,  \nonumber \\
Dos &=&2\ Dol\,\text{,}  \label{simulation_num}
\end{eqnarray}
implying 
\begin{equation}
r_{E}=\sqrt{2} \cdot 10^{4}\frac{G_{N}M}{c^{2}}\simeq \left\{ 
\begin{array}{l}
\left. 
\begin{array}{l}
2 \cdot 10^{18}\ \text{m for }M=10^{11}M_{Sun}
\end{array}
\right\} \text{\ a galaxy} \\ 
\left. 
\begin{array}{l}
2 \cdot 10^{20}\ \text{m for }M=10^{13}M_{Sun} \\ 
2 \cdot 10^{21}\ \text{m for }M=10^{14}M_{Sun} \\ 
2 \cdot 10^{22}\ \text{m for }M=10^{15}M_{Sun}
\end{array}
\right\} \text{ a cluster}\,.
\end{array}
\right.   \label{radius_Einstein_ring_mirage_simulation}
\end{equation}
Distance scales involved in gravitational mirages prevent us from using the weak field 
approximation with our bounds (\ref{estimated_value_of_gamma_Weyl_Cassini}) 
on a positive $\gamma _{W}$, but a negative value of the parameter can be considered.

Whichever type of lensing event (microlensing or gravitational mirages) we consider, 
conditions for light deflection to take place ($r_{0}<r_{null}$, $\cite{Pireaux 2003
critical distances in Weyl}$) are fulfilled each time we work in the weak
field limit, according to the coincidence of $r_{weak\ field}$ (\ref
{weak_field_radius_Weyl}) and $r_{null}$ (\ref{r_null_Weyl}).

\subsection{Constraints on a negative $\gamma _{W}$}

\qquad With the above remarks in mind, we now extract 
information from microlensing events or gravitational mirages for a 
negative value of the linear Weyl parameter.\medskip

\subsubsection{Asymptotic weak field light deflection angle}

\quad \newline
In the weak field limit, using $r_{0}\simeq
Dol\ \sin \vartheta _{I}$ at zero order in $V_{W}(r)/c^{2}$, 
Equation (\ref{angle_deflex_weak_above_parsec_WEYL}) 
can be rewritten, at first order, as 
\begin{equation}
\hat{\alpha}_{\text{weak field}}(\vartheta _{I},Dol)\simeq +\frac{4\beta _{W}%
}{Dol\ \sin \vartheta _{I}}-\gamma _{W}\ \ Dol\ \sin \vartheta
_{I}\,\smallskip  \label{angle_deflex_weak_above_parsec_fct_theta_WEYL}
\end{equation}
where $\vartheta _{I}$ is the angular position of the observed image (I),
with respect to the O-L axis (see Figure \ref{plan_deflex_all}). 
The Weyl deflection angle in the weak field limit 
cannot be rescaled to the Einsteinian 
prediction by a redefinition of the mass, as it is the case in Tensor Scalar theories.
So, predictions of Weyl gravity are expected to be qualitatively
different from the general relativistic ones.\medskip

\subsubsection{Lens equation in the weak field limit}

\quad \newline
The lens equation associated with a converging deflection angle (\ref
{angle_deflex_weak_above_parsec_fct_theta_WEYL}) in the weak field limit and
small angle approximation ($\vartheta _{I}$, $\vartheta _{S}$, $\hat{\alpha}%
\ll \sqrt{3}$ rad) is 
\begin{equation}
\overrightarrow{\vartheta _{I}}^{2}-\frac{\overrightarrow{\vartheta _{S}}}{%
1+n_{W}}\overrightarrow{\ \vartheta _{I}}-\ \vartheta _{W}^{2}=0\,,
\label{posit_ima_petit_a_Weyl}
\end{equation}
where we have defined 
\begin{eqnarray}
\smallskip   \nonumber \\
&&\vartheta _{E}
\begin{array}[t]{l}
\equiv \text{ angular radius of the Einstein ring} \\ 
\equiv \sqrt{\frac{4G_{N}M}{c^{2}}\frac{Dls}{Dol\ Dos}}\,,
\end{array}
\smallskip   \label{Einstein_angle} \\
&&\vartheta _{W}
\begin{array}[t]{l}
\equiv \text{ angular radius of the Weyl ring} \\ 
\equiv \frac{1}{\sqrt{1+n_{W}}}\vartheta _{E}\,,
\end{array}
\smallskip   \label{Weyl_angle} \\
&&r_{E}
\begin{array}[t]{l}
\equiv \text{radius of the Einstein ring} \\ 
\simeq \vartheta _{E}\ Dol\,,
\end{array}
\smallskip   \label{Einstein_radius} \\
&&r_{W}
\begin{array}[t]{l}
\equiv \text{radius of the Weyl ring} \\ 
\simeq \vartheta _{W}\ Dol\,,
\end{array}
\smallskip   \label{weyl_radius} \\
&&n_{W}
\begin{array}{l}
\equiv \gamma _{W}\ \frac{Dls\ Dol}{Dos}\,\text{.}
\end{array}
\label{Weyl_dimensionless_parameter}
\end{eqnarray}
\newline
For a given O-L-S configuration, Equation (\ref{posit_ima_petit_a_Weyl}) 
allows to compute the position of images, or inversely, to infer the position 
of the light source from the observed position of images.\newline
The Weyl lens equation is quadratic, like in General Relativity. 
The corrective factor with respect to General Relativity, 
$1/(1+n_{W})$, reduces to $1$ when $\left. \gamma _{W}\right| _{GR}=0$ 
and is greater than 1 for a negative Weyl parameter. Indeed,this lens equation 
is only valid in the weak field limit (\ref{weak_field_radius_Weyl}), 
leading to $n_{W}$ always smaller in norm than $1$. \newline 
Changing the O-L-S distances in the microlensing or
gravitational mirage models (\ref{simulation_num_microlens} or \ref
{simulation_num}) would change the Einstein/Weyl radii (\ref{Einstein_angle}%
, \ref{Einstein_radius}, \ref{radius_Einstein_ring_microlensing_simulation}; 
\ref{Weyl_angle}, \ref{weyl_radius}, \ref{radius_Einstein_ring_mirage_simulation}), 
but the radius at which the geodesic potential for photons cancels
(\ref{r_null_Weyl}), the interesting closest approach distance (\ref
{r0_0_Weyl}) and the weak/strong field limiting radii (\ref
{weak_field_radius_Weyl}, \ref{strong_field_radius_Weyl}) would remain
unchanged.\newline
Interestingly, the predictions that we derive from Equation 
(\ref{posit_ima_petit_a_Weyl}) not only depend on the parameter of the Weyl 
theory through $n_{W}$, but also on the physical properties of the O-L-S system 
via a combination of the distances $Dls$, $Dol$, and$\ Dos$.
\medskip

\FRAME{fpFU}{10.4757cm}{20.3254cm}{0pt}{\Qcb{Usual thin lens model of a
gravitational mirage where O is the observer; L, the gravitational lens; I,
the image formed; $\overrightarrow{\widehat{\alpha }}$, the asymptotic light
deflection angle; $\overrightarrow{\overline{b}}$, the impact parameter of
the lens on the O-S direction; $\overrightarrow{\vartheta _{I}}$, the
angular position of the observed image, with respect to the O-L axis; $%
\overrightarrow{\vartheta _{S}}$, the angular position of the light source,
with respect to the O-L axis. In the case of a diverging lens,
there is no image formed from a divergent ray on the opposite side of the source 
with respect to the O-L axis.} }{\Qlb{%
plan_deflex_all}}{plan_deflex_all.eps}{\special{language "Scientific
Word";type "GRAPHIC";display "ICON";valid_file "F";width 10.4757cm;height
20.3254cm;depth 0pt;original-width 0pt;original-height 0pt;cropleft
"0";croptop "1";cropright "1";cropbottom "0";filename
'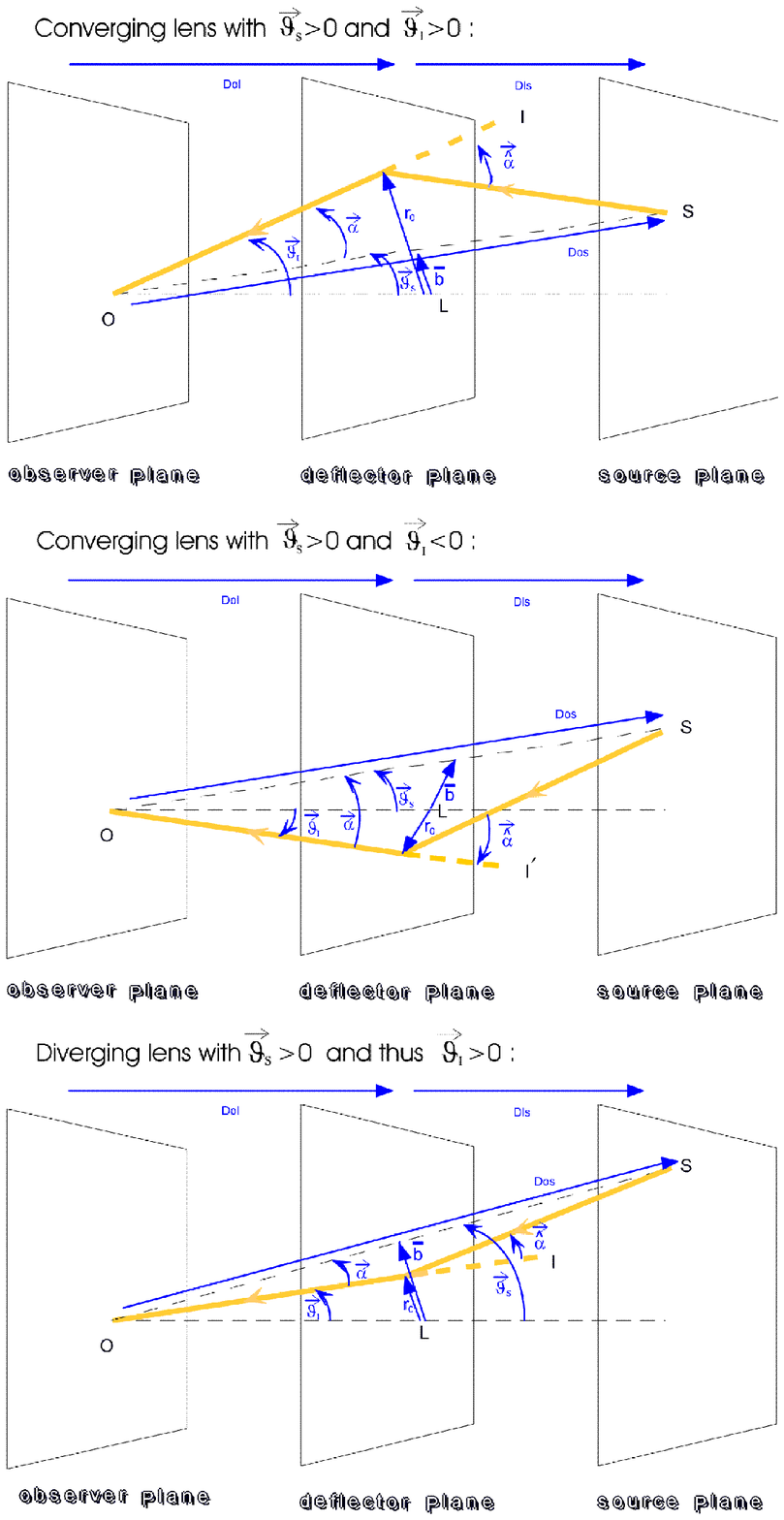';file-properties "XNPEU";}%
}

\underline{In case of alignment ($\vartheta _{S}=0$)} of the observer, the
lens and the source, the image is a ring of angular radius $\vartheta _{I}=\vartheta _{W}$.
The predicted width of the ring ($2\ d\vartheta _{I}$) given as a function
of the angular diameter of the light source ($d\vartheta _{S}$) \emph{will
be different from the Einsteinian one}, even if we refer to the natural size of the
physical problem ($\vartheta _{W}$ instead of $\vartheta _{E}$ used in General Relativity): 
\[
d\vartheta _{I}=\frac{1}{2(1+n_{W})}d\vartheta _{S}\,\text{.}\medskip 
\]
The Weyl ring will also be thicker than its general relativistic
counterpart. Consequently, the amplification, $\mu $, will be larger: 
\[
\mu \simeq \frac{2\ \vartheta _{W}}{\left( 1+n_{W}\right) d\vartheta _{S}}%
\,, 
\]
and infinite in the limit of a point-like lens mass, as in General Relativity.\medskip

\underline{In case of misalignment ($\vartheta _{S}>0$)}, the ring breaks up
into two arcs located at 
\[
\vartheta _{I+/-}=\frac{\vartheta _{S}\pm \sqrt{\vartheta _{S}^{2}+4\
\vartheta _{W}^{2}\ \left( 1+n_{W}\right) ^{2}}}{2\left( 1+n_{W}\right) }\,%
\text{,} 
\]
and separated by an angle 
\[
\Delta \vartheta _{I}\equiv \vartheta _{I+}-\vartheta _{I-}=\frac{\sqrt{%
\vartheta _{S}^{2}+4\ \vartheta _{W}^{2}\ \left( 1+n_{W}\right) ^{2}}}{%
\left( 1+n_{W}\right) }\text{\thinspace .\smallskip } 
\]
The sign $+/-$ for $\vartheta _{I}$ distinguishes between images formed on the 
same side or on the oposite side of the light source with respect to the O-L 
axis.\newline
The width of the arcs ($2d\vartheta _{I\pm }$) is given by 
\[
d\vartheta _{I\pm }=\frac{1}{2\left( 1+n_{W}\right) }\left[ 1\pm \frac{1}{%
\sqrt{1+4\ \left( \vartheta _{W}/\vartheta _{S}\right) ^{2}\ \left(
1+n_{W}\right) ^{2}}}\right] \ d\vartheta _{S}\text{\thinspace .} 
\]
Consequently, each image will be more amplified than in General Relativity,
with the corresponding amplifications 
\begin{eqnarray*}
\mu _{+/-} &\simeq &\frac{1}{4\left( 1+n_{W}\right) ^{2}}\quad \left[ 2\pm
\left( \left( 1+n_{W}\right) \frac{\Delta \vartheta _{I}}{\vartheta _{S}}+%
\frac{1}{\left( 1+n_{W}\right) }\frac{\vartheta _{S}}{\Delta \vartheta _{I}}%
\right) \right] \bigskip \\
&=&\frac{1}{\left( 1+n_{W}\right) ^{2}}\quad \frac{\pm \left[ \overline{B}_{%
\widehat{W}}\pm \sqrt{\overline{B}_{\widehat{W}}^{2}+4}\right] ^{2}}{4%
\overline{B}_{\widehat{W}}\ \sqrt{\overline{B}_{\widehat{W}}^{2}+4}}\,,
\end{eqnarray*}
\medskip where we define the following dimensionless quantities with respect
to $\overline{b}$, the impact parameter of the deflector on the O-S
direction: 
\begin{eqnarray*}
\overline{B}_{E} &\equiv &\overline{b}/r_{E}\,,\text{ dimensionless
Einstein impact parameter of the deflector,} \\
\overline{B}_{W} &\equiv &\overline{b}/r_{W}\,,\text{ dimensionless Weyl
impact parameter of the deflector,} \\
\overline{B}_{\widehat{W}} &\equiv &\frac{1}{\left( 1+n_{W}\right) }%
\overline{B}_{W}\text{\thinspace ,} 
\begin{array}[t]{l}
\text{ normalized dimensionless Weyl impact} \\ 
\text{parameter of the deflector.}
\end{array}
\end{eqnarray*}
\newline
Hence, the total amplification, when the two images are not resolved, is
again larger than the Einsteinian value: 
\begin{equation}
\mu _{tot}\simeq \frac{1}{\left( 1+n_{W}\right) ^{2}}\quad \frac{\overline{B}%
_{\widehat{W}}^{2}+2}{\overline{B}_{\widehat{W}}\sqrt{\overline{B}_{\widehat{%
W}}^{2}+4}}\,\text{.}
\label{total_amplification_microlensing_Weyl} 
\end{equation}

\subsubsection{Microlensing}

\quad \newline
The above expression and 
\[
\overline{B}_{\widehat{W}}(t)\simeq \overline{B}_{\widehat{W}\text{ }%
0}\left[ 1+\frac{t^{2}}{T_{0}^{2}}\right] 
\]
are the relevant equations to be used for microlensing amplification curves. 
$T_{0}$ is the time correponding to the minimal normalized dimensionless
Weyl impact parameter of the deflector, $\overline{B}_{\widehat{W}\text{ }0}$%
.\newline
Strictly speaking, we cannot rescale the amplification curve ($%
\overline{B}_{\widehat{W}}\mapsto \overline{B}_{E}$) to fit Einsteinian predictions
because of the front factor in (\ref{total_amplification_microlensing_Weyl}). 
In view of the upper bound for a negative
linear parameter (\ref{estimated_value_of_gamma_negative_Weyl}) and of the
typical distances for microlensing events (\ref{simulation_num_microlens}),
we find an upper bound for the corrective factor 
\begin{equation}
\left. \frac{1}{\left( 1+n_{W}\right) }\right| _{\text{{\tiny %
microlens (\ref{simulation_num_microlens})}}}-1 \stackrel{\medskip {\tiny (}%
\text{{\tiny \ref{estimated_value_of_gamma_negative_Weyl})}}}{\lesssim }10^{-11}\text{ .}
\label{corrective_factor_negative_microlensing_Weyl}
\end{equation}
This corrective factor might be very small for microlensing
events if $\left| \gamma _{W}\right| $ happens to be much smaller than our
present upper bound on $\gamma _{W}$ (\ref{estimated_value_of_gamma_negative_Weyl}).

The optical depth of microlensing $\cite{Narayan 1997 lectures gravitational
lensing}$ is the probability, at a given time, that a light source be within
the corresponding ring radius of a given star lens; hence, it is the
probability for this light source to be lensed. To estimate the optical
depth in the setting of the Weyl theory, we must integrate over the surface
included in the ring radius, thus, over $\pi \vartheta _{W}^{2}$. According
to (\ref{Weyl_angle}), the corrective factor 
(\ref{corrective_factor_negative_microlensing_Weyl}) enters in the
integral. One might have hoped that the Weyl theory would
substantially increase the optical depth, so maybe to account for 
observations leading to a larger value than initially estimated with General
Relativity. Unfortunately, owing to the upper bound on $\left| \gamma
_{W}\right| $ in (\ref{estimated_value_of_gamma_negative_Weyl}), the
correction is irrelevant with respect to observational uncertainties.

\subsubsection{Gravitational mirages}

\quad \newline
For distance scales involved in gravitational mirages, our constraint 
(\ref{estimated_value_of_gamma_negative_Weyl}) 
on a negative Weyl parameter leads to the following maximum deviation from 
General Relativity:
\begin{equation}
\left. \frac{1}{\left( 1+n_{W}\right) }\right| _{\text{{\tiny %
mirages (\ref{simulation_num})}}}-1 \stackrel{\medskip {\tiny (}%
\text{{\tiny \ref{estimated_value_of_gamma_negative_Weyl})}}}{\lesssim }10^{-5}\text{ .}
\label{corrective_factor_negative_mirages_Weyl}
\end{equation}
This might be negligible. As gravitational lenses are always convergent for a 
negative $\gamma_{W}$, the number and parity (position of $\vartheta _{I}$ relative to 
$\vartheta _{S}$ on the diagram) of the images formed in Weyl gravity 
is analogous to the Einsteinian case. Consequently, deviations from General Relativity 
are only quantitative, and solely a statistic of gravitational lensing events with 
different lens mass distribution models would settle the question whether the corrective 
factor might have observable consequences or not.

\subsection{Testing the Mannheim-Kazanas parametrization ($\gamma _{W}>0$)}

\qquad We have argued, in Paragraph \ref{discussion_pertinence_weakfield_limit_W} 
that we could not so far use the weak field limit to further constrain 
a positive \textbf{$\gamma _{W}$}. However, we can investigate the possibility to 
invalidate the Mannheim-Kazanas theory (\ref{parametrization_Mannheim_Kazanas}, 
\ref{definition_Mannheim_theory}),
or try to extract predictions on microlensing or gravitational
mirages from this theory where the weak field limit (\ref
{weak_field_radius_Weyl}) extends to cosmological distances.

\subsubsection{Lens equation in the weak field limit}

\quad \newline
Accordingly, we use the weak field lens equation (\ref
{posit_ima_petit_a_Weyl}) with the corresponding definitions (\ref
{Einstein_angle}, \ref{Weyl_angle}, \ref{Einstein_radius}, \ref{weyl_radius}%
, \ref{Weyl_dimensionless_parameter}). Nevertheless, if assuming a positive
value of the linear Weyl parameter, additional conditions are necessary
when the deflection angle is divergent [$\overrightarrow{\widehat{\alpha }}%
<0 $ for $\vartheta _{I}>\vartheta _{I0}=\arcsin \left\{ \frac{r_{0\ 0}}{Dol%
}\right\} \,$, where $r_{0\ 0}$ is defined in Equation (\ref{r0_0_Weyl})]. 
Those conditions might be infered from Figure \ref{plan_deflex_all}:
\begin{equation}
\left. 
\begin{tabular}{l}
sign$(\overrightarrow{\vartheta _{I}})=$sign$(\overrightarrow{\vartheta _{S}}%
)\,\smallskip $ \\ 
$\vartheta _{S}\neq 0\,\smallskip $ \\ 
$\vartheta _{S}>\vartheta _{I}\,$%
\end{tabular}
\right\}  \label{posit_ima_petit_a_additional_cdt_Weyl}
\end{equation}
The corrective factor corresponding to the Mannheim-Kazanas parametrization, 
$1/(1+n_{W})$, is still positive but smaller than 1.

\subsubsection{Microlensing}

\quad \newline
As far as microlensing is concerned, the interesting closest approach
distance $r_{0\ 0}$ (\ref{r0_0_Weyl}), at which light deflection becomes
divergent, is larger than the microlensing Einstein radius (\ref
{radius_Einstein_ring_microlensing_simulation}) and hence is irrelevant.
Thus, light deflection is always convergent on microlensing scales,
which means no qualitative deviations from General Relativity. Moreover, an
estimation of the order of magnitude of the corrective factor for the
Mannheim-Kazanas parametrization, 
\begin{equation}
\left. 1-\frac{1}{1+n_{W}}\right| _{\text{{\tiny microlens (\ref
{simulation_num_microlens})}}}\stackrel{\text{{\tiny (\ref
{parametrization_Mannheim_Kazanas})}}}{\simeq} 10^{-5}\,, 
\label{corrective_factor_microlensing_Mannheim_Weyl}
\end{equation}
shows that microlenses are not an appropriate tool to test the
Mannheim-Kazanas theory.\medskip

\subsubsection{Gravitational mirages}

\quad \newline
Regarding\emph{\ }gravitational mirages, $r_{0\ 0}$ (\ref{r0_0_Weyl}) 
is relevant ($\vartheta _{I0}< \vartheta _{E}$) 
to the most massive clusters ($M\sim 10^{15}M_{Sun}$). 
But in this case, the O and L points of the photon trajectory
lie about on the edge of the weak field limit corresponding to the Mannheim-
Kazanas parametrization, while $r_{0}\sim r_{E}$ is still well within it.\newline
Nevertheless, let us discuss the corresponding image-position diagram
(Figure \ref{lens_weyl_mannheim}) which represents the image position ($%
\vartheta _{I}$) solution to the lens equation (\ref{posit_ima_petit_a_Weyl}%
) as the crossing of two curves for a given source position ($\vartheta _{S}$%
): 
\begin{eqnarray*}
F_{1}(\overrightarrow{\vartheta _{I}}) &\equiv &\overrightarrow{\vartheta
_{I}}-\overrightarrow{\vartheta _{S}} \\
F_{2}(\overrightarrow{\vartheta _{I}}) &\equiv &sign(\vartheta _{I})\cdot
\left[ \left( \frac{1}{1+n_{W}}-1\right) \overrightarrow{\vartheta _{S}}+%
\frac{\vartheta _{W}^{2}}{\overrightarrow{\vartheta _{I}}}\right] 
\end{eqnarray*}
\newline
The $F_{2}$-curve crosses the $\vartheta _{I}$-axis at a value $\vartheta
_{I0}$ corresponding to $r_{0\text{ }0}$, separating the divergent from the
convergent contribution. $F_{2}$ presents no foldings. This means that the
predictions of the Mannheim-Kazanas theory will be the same as the Einsteinian ones
from the point of view of the number of images and their parity, which seems to
prevent this theory to be tested using gravitational mirages. One can just say that
the Mannheim-Kazanas gravity even needs more dark matter than General
Relativity (the Weyl radius being smaller than the Einstein ring), because
the corrective factor is smaller than $1$. To illustrate this, note that the
simulation discussed (\ref{simulation_num}) provides 
\begin{equation}
\fl\left. \frac{1}{1+n_{W}}\right| _{\text{{\tiny mirages (\ref
{simulation_num})}}}\stackrel{\text{{\tiny (\ref
{parametrization_Mannheim_Kazanas})}}}{\simeq }\frac{1}{1+0.75 \cdot 10^{-15+x}}%
\text{\quad where }M=10^{x}M_{Sun}
\label{corrective_factor_gravlens_Mannheim_Weyl}
\end{equation}
and clusters with a mass of $\sim 10^{15}M_{Sun}$
are on the edge of the weak field approximation, as explained.
\newline
Note that when $\vartheta _{S}< \vartheta _{I0}$, the two images formed in case 
of misalignment are due to two convergent rays; when $\vartheta _{S}= \vartheta _{I0}$, 
one image is from a convergent ray whereas the other one is from a nondeflected ray of light; 
and when $\vartheta _{S}>\vartheta _{I0}$, one image is from a convergent ray and the other 
one is from a divergent ray.

\FRAME{fhF}{14.0342cm}{10.5416cm}{0pt}{\Qcb{Image-position diagram ($F_{1}$-
and $F_{2}$-curves) for a point-like mass lens model ($M_{cluster}=10^{15}M_{Sun}$) 
in the framework of the Mannheim-Kazanas theory ($\gamma_{W}\simeq 10^{-26}\ $m$^{-1}$).
\newline
The crossings between the $F_{1}$- and $F_{2}$-curves correspond to the
image positions ($\vartheta _{I}$). 
The case represented here is for $\overrightarrow{\vartheta_{S}}>0$ and 
$F_{1}$ is plotted at $\vartheta _{S}=$0 (alignment), +10
arcsec and +30 arcsec. The values of the variables in the simulation are
given in the text.}}{\Qlb{lens_weyl_mannheim}}{lens_weyl_mannheimbis.eps}{%
\special{language "Scientific Word";type
"GRAPHIC";maintain-aspect-ratio TRUE;display "ICON";valid_file "F";width
14.0342cm;height 10.5416cm;depth 0pt;original-width 0pt;original-height
0pt;cropleft "0";croptop "1";cropright "1";cropbottom "0";filename
'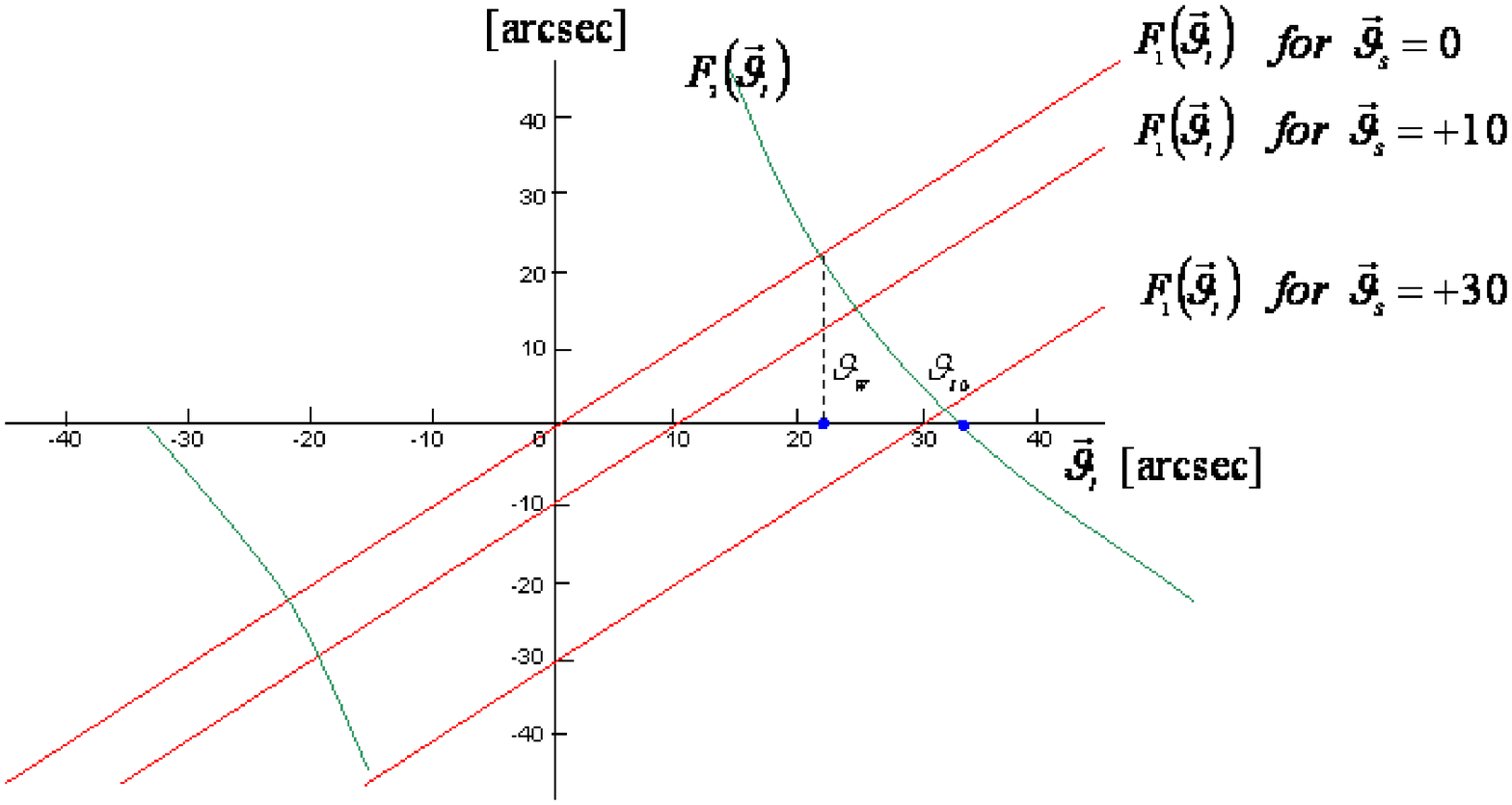';file-properties "XNPEU";}%
}

\subsection{Attempts in the literature to deduce a negative value for $%
\gamma _{W}$}

\qquad In article $\cite{Edery 1998 lensing in Weyl theory}$, Edery and Paranjape are
interested in a negative value of the linear Weyl parameter in order to
explain gravitational mirages without dark matter. They extract an order of
magnitude for $\gamma _{W}$ from observations of giant arcs in clusters. It
happens to be just the same order of magnitude (the inverse of the Hubble
length) as needed in the\ Mannheim-Kazanas parametrization (\ref
{parametrization_Mannheim_Kazanas}) but with the opposite sign.\newline
The idea behind this estimation is to equate the predicted Einstein radius (%
\ref{Einstein_radius}) based on an estimated value of the total
gravitational mass (luminous plus an \emph{ad hoc} amount of dark matter, to
fit observations in the framework of General Relativity) with the Weyl radius 
(\ref{weyl_radius}) for the same O-L-S system, but this time with only the
luminous matter.\newline

In the Cluster lenses used for this purpose, namely A 370, A 2390 and Cl
2244-02, the ratios of the luminous over the total mass were estimated
respectively to be $M_{L}/M_{tot}\sim 1/200,\ \sim 1/120$ and $<1/100$,
thanks to a complete modeling of each lens in the setting of General Relativity $\cite
{Bergmann 1990 Lensing arcs clusters},\,\cite{Grossman 1989 Lensing arcs
clusters},$ $\cite{Pello 1991 Lensing arcs clusters}$.\medskip
However, it is the general relativistic estimate (not a Weyl estimate) of
the luminous mass that the authors implicitly inserted into the Weyl radius 
when using the above luminous to total mass ratios.\medskip

Also, in their estimation of the Weyl radii, the authors use the general
relativistic relation between the redshift and the distance. In the
framework of General Relativity, it is the Robertson-Walker metric, solution
to the Einstein equations in presence of matter, which is used to infer the
concept of cosmological distance as a function of the redshift, the matter
density and the curvature of the universe. However, in the Weyl theory, the
Robertson-Walker metric is \emph{not} a solution of the Bach equations with
matter! Indeed, the Weyl tensor associated with the Robertson-Walker metric 
is null, hence the Bach tensor ($B_{\mu \nu }$, Equation (2) in 
$\cite{Pireaux 2003 critical distances in Weyl}$) function of the Weyl tensor 
is null too, which means that the Bach equations \cite{Bach 1921 Bach equation} 
\emph{in presence of}
matter, 
\[
B_{\mu \nu }=-1/4\ \sqrt{\kappa }\ T_{m\ \mu \nu }\,,\medskip 
\]
are not satisfied!\newline
The concept of distance in the Weyl theory
(as well as the concept of time) requires the specification of the conformal
factor. This conformal factor is crucial in cosmology to obtain the
appropriate Weyl prescription relating observed redshifts to 
cosmological distances. Otherwise, one can only work in terms of distance
ratios (or time ratios, or the mixed time to distance ratios) and angles.%
\newline
Note that, in the limit of small redshifts, the rough relationship between 
redshift and distance given in (\ref
{rough_estimate_redshift_from_distance}) together with the approximation of
an Euclidean space ($Dls=Dos-Dol$) are sufficient to recover the same order
of magnitude for $\gamma _{W}$ as the one calculated by Edery \emph{et al}. $%
\cite{Edery 1998 lensing in Weyl theory}$. We obtain as well the natural
connection between $\gamma _{W}$ and the inverse of the Hubble length: 
\begin{equation}
\begin{array}{lcl}
\left| \gamma _{W}\right| & = & \left( 1-\frac{M_{L}}{M_{tot}}\right) \frac{%
Dos}{Dls\ Dol} \\ 
\smallskip & \stackrel{{\tiny (}\text{{\tiny \ref
{rough_estimate_redshift_from_distance}) and Euclidean approximation}}}{%
\simeq } & \left( 1-\frac{M_{L}}{M_{tot}}\right) \frac{z_{s}}{(z_{s}-z_{l})\
z_{l}}\frac{H_{o}}{c} \\ 
& \stackrel{\text{{\tiny A370, A2390 and Cl2244-02}}}{\sim } & 5
\cdot 10^{-26}-6 \cdot 10^{-26}\ \text{m}^{-1}\,.\smallskip
\end{array}
\label{estimate_gamma_negative_Edery_Weyl}
\end{equation}
The approximation (\ref{rough_estimate_redshift_from_distance}) can be
obtained simply from the observational redshift which is related to the
recession speed of galaxies ($v_{recess}$) by a simple Doppler effect, and
from the empirical Hubble law based on observations: 
\[
\frac{v_{recess}}{cst\ D}\simeq 1\quad \text{with }cst=\left. H_{0}\,\right|
_{exp}\,. 
\]
Interestingly, the Hubble law cited above, and the measured redshift (which
is a ratio of frequencies) do not require to specify the conformal factor in
the Weyl line element.\medskip

Another point concerns \textbf{the weak field approximation} that the
authors Edery and Paranjape use, through lens equation (\ref
{posit_ima_petit_a_Weyl}) and the corresponding definitions of the Weyl
radius (\ref{weyl_radius}).\newline
The gravitational mirages considered in their article, 
\[
\begin{tabular}{c|c|c}
lens & $z_{L}$ & $z_{S}$ \\ \hline
$A$ $370$ & $0.375$ & $0.724$ \\ 
$A$ $2390$ & $0.231$ & $0.913$ \\ 
$Cl$ $2244-02$ & $0.331$ & $0.83$%
\end{tabular}
\]
correspond to observer-lens or to lens-source distances, as can be
calculated using the rough estimation (\ref
{rough_estimate_redshift_from_distance}), that are dangerously close to the
weak field limit associated with their estimated value of $\gamma _{W}$. This
leaves us dubious about the above result (\ref
{estimate_gamma_negative_Edery_Weyl}).\medskip

Finally, in another article $\cite{Edery 2000 weyl gauge choice}$, Edery 
\emph{et al}. speak of the ``theoretical arbitrariness'' of the choice of
the conformal factor in the Weyl spherically symmetric solution (Equation
(3) in $\cite{Pireaux 2003 critical distances in Weyl}$). They furthermore
use the conformal transformation given in (\ref
{conform_transfo_change_sign_Weyl}) to argue that the parameter $\gamma _{W}$
might be measured as positive for matter particles with $ds_{matter}^{2}%
\equiv ds^{2}(r^{^{\prime }},\gamma _{W}^{^{\prime }}>0)$ in (\ref
{conform_transfo_change_sign_Weyl}); and on the contrary, as negative for
photons with $ds_{photons}^{2}\equiv ds^{2}(r,\gamma _{W}<0)$ in (\ref
{conform_transfo_change_sign_Weyl}). Their aim is to try to connect their
estimate of $\gamma _{W}$ (\ref{estimate_gamma_negative_Edery_Weyl}) to the 
Mannheim-Kazanas theory (\ref{definition_Mannheim_theory}) and explain the
discrepancy in the sign of the linear Weyl parameter.\newline
However, if the motion of ultra-relativistic particles and photons does not
depend on the conformal factor $\chi ^{2}(r)$ and the choice of $\chi
^{2}(r)$ is arbitrary when restricting to those types of particles, we have
shown that on the contrary, the motion of matter $\cite{Pireaux 2003 critical
distances in Weyl}$ as well as the definition of
distance scales and timescales crucially depends on it. Because our world is
made of matter, it is not conformally invariant (we indeed have clocks and
rods to make measurements). Hence, the conformal factor is not
arbitrary, as long as we are looking for a theory to describe Nature. Nature
has chosen a specific conformal factor. The Weyl
theory should even be called ``Weyl theories'', because it in fact corresponds to
a class of theories, each theory being specified by a choice of $\chi
^{2}(r)$ and ($\gamma _{W}$, $k_{W}$) while $\beta _{W}\equiv \frac{G_{N}M}{%
c^{2}}$. The Mannheim-Kazanas theory is a particular example.\newline
Moreover, even though the linear $\gamma _{W}$-contribution to the effective
potential might have an opposite effect on massless in comparison with
massive particles (Subsection 3.1 in $\cite{Pireaux 2003 critical distances
in Weyl}$), it is necessary to use the same radial variable in
order to have the same definition of $\gamma _{W}$ when comparing estimates 
of $\gamma _{W}$ obtained from photon trajectories to those obtained from massive 
particle motion.

\section{Conclusions}

\qquad Regarding the change of the apparent position of stars, the
weak field regime applied to the VLBI data gave an upper bound on the
linear parameter: $\left| \gamma _{W}\right| \sim 10^{-18}\ $m$^{-1}$ (\ref
{estimated_value_of_gamma_Weyl}) that was improved with the use of the
recent Cassini Doppler Data, to lead to $\left| \gamma _{W}\right| \sim
10^{-19}\ $m$^{-1}$ (\ref{estimated_value_of_gamma_Weyl_Cassini}). However,
Solar System experiments do not settle the sign of $\gamma _{W}$. 
A positive $\gamma _{W}$ decreases the light deflection angle, while a
negative one increases it with respect to General Relativity. 
A negative parameter might
thus be an alternative to a too large amount of dark matter. Analyzing the
expression for the asymptotic deflection angle in the Weyl theory, it was
found to allow for a diverging effect at closest approach distances larger
than $r_{0\ 0}$ (\ref{r0_0_Weyl}) function of the mass of the lens, \emph{if 
}$\gamma _{W}$ \emph{is positive}. However, the convergent to divergent
transition is not explicitly relevant to Solar System experiments.\medskip 

We considered, when applicable, the weak field approximation to study
microlensing and gravitational mirages. The strong field limit was found to
be useless for our purposes; because we only have in hand an upper bound on $%
\left| \gamma _{W}\right| $. The lower bound given by
General Relativity ($\gamma _{W}=0$) leads to a strong field limit
only valid at infinity.

\underline{For a negative\emph{\ }$\gamma _{W}$}, the condition for unbound
photon orbits derived in the preceeding article $\cite{Pireaux 2003 critical
distances in Weyl}$ and the existence of gravitational mirages were
used to improve constraints on the $\gamma _{W}$-parameter obtained from Solar 
System data. The upper bound on the absolute value of $\gamma _{W}
$ was accordingly lowered to about $10^{-31}\ $m$^{-1}$.\newline
The characteristics of the microlensing or gravitational mirage curve 
in the Weyl theory cannot, by a simple rescaling of the mass or the ring radius, 
be recast into the Einsteinian predictions. 
However, the corrective factor, $1/(1+n_{W})$, function of $%
\gamma _{W}$ and the O-L-S distances, is small. Indeed, it is equal to $1$
when $\gamma _{W}$ is equal to zero (General Relativity), and differs from its 
Einsteinian value with a maximum of $\sim +10^{-11}$ for microlenses, or 
$\sim +10^{-5}$ for gravitational mirages. 
This means that it is negligible for microlenses and might effectively be negligible 
for gravitational mirages too. The latter point requires further study. 
Our estimate of the corrective factor is based upon our upper bound on 
$\left| \gamma _{W}\right| $, for a negative Weyl parameter and a point mass lens model.

\underline{In the Mannheim-Kazanas theory ($\gamma _{W}\simeq 10^{-26}\ $m$%
^{-1}>0$)}, gravitational mirages do not seem to be an appropriate test.
Indeed, the predictions will be the same as the general relativistic ones,
from the point of view of the number of images and of their parity. One can
just say that the Mannheim-Kazanas gravity needs even more dark matter than
General Relativity, because the Weyl radius is smaller than the Einstein one.\newline
In microlensing, the interesting closest approach distance separating the
convergent from the divergent contributions ($r_{0\ 0}$) is cosmological in
the Mannheim-Kazanas theory, and thus irrelevant. Moreover, the smallness of
the corrective factor $\sim (1-10^{-5})$ showed that one cannot
distinguish between General Relativity and the Mannheim-Kazanas theory from
the point of view of microlensing, either.\medskip

We commented on the concept of distance that needs to be defined
consistently in the Weyl theory. Also, to be
consistent, the weak (or strong) field limit on the radial
distance, measured from the gravitational deflector, has to be verified on
the photon path, all the way from the light source to the observer. Thus,
the limit does not only apply to the lens-observer and lens-source distances, 
but also to the closest approach distance of the photon onto the lens.\medskip

Even though the bounds we obtained on $\gamma _{W}$ are conservative in
comparison with Mannheim's estimated value of $\gamma _{W}\sim +10^{-26}\ $m$%
^{-1}$, they are preferable because they are not biased by any arbitrary
assumption on the conformal factor. Moreover, the particular value of
Mannheim and Kazanas belongs to the allowed range that we derived.


\ack{%
The research work presented here was carried out in the FYMA
Institute at the University of Louvain la Neuve, during a Ph.D.thesis
financed by the I.I.S.N research assistantship. We are grateful to Professor 
M. Festou (LAT, Observatoire Midi-Pyr\'{e}n\'{e}es, Toulouse) for pertinent advice.}


\bigskip

\begin{thebibliography}{Tury et al.2003}
\bibitem[Bach 1921]{Bach 1921 Bach equation}  R. 
Bach. \textit{Zur Weylschen Relativit\"{a}tstheorie und der Weylschen 
Erweiterung des Kr\"{u}mmungstensorbegriffs.}. Mathematische Zeitschrift 9, 
110-135 (1921).

\bibitem[Be et al.1990]{Bergmann 1990 Lensing arcs clusters}  A. G.
Bergmann, V. Petrosian and R. Lynds. \textit{Gravitational lens models of
arcs in clusters}. Astrophysical Journal, 350, 23-35 (1990).

\bibitem[Be et al.2003]{Bertotti 2003 Cassini gamma measurement}  B.
Bertotti, L. Iess and P. Tortora. \textit{A test of General Relativity using
radio links with the Cassini spacecraft.} Letters to Nature, 425, 374-376
(2003).

\bibitem[Co et al.1974]{Counselman 1974 VLBI gamma determination}  
C.C. Counselman III, S.M. Kent, C.A. Knight, I.I. Shapiro, T.A. Clark, 
H.F. Hinteregger, A.E.E. Rogers and A.R. Whitney. \textit{Solar gravitational 
deflection of radio waves measured by {Very-Long-Baseline Interferometry}.} 
Physical Review Letter, 33, 1621-1623 (1974).

\bibitem[EdPa1998]{Edery 1998 lensing in Weyl theory}  A. Edery and M. B.
Paranjape. \textit{Classical Tests for Weyl Gravity: Deflection of Light and
time Delay}. Physical Review D., 58, 1-8 (1998) astro-ph/9708233.

\bibitem[Ed et al.2001]{Edery 2000 weyl gauge choice}  A. Edery, A. A. M\'{e}%
thot and M. B. Paranjape. \textit{Gauge choice and geodetic deflection in
conformal gravity}. General Relativity and Gravitation, 33, 2075-2079 (2001), 
astro-ph/0006173.

\bibitem[Eu et al.1999]{Eubanks 1999 advances in solar system tests}  T. M.
Eubanks, J. O. Martin, B. A. Archinal, F. J. Josties, S. A. Klioner, S.
Shapiro and I. I. Shapiro. \textit{Advances in Solar System tests of gravity}%
. \TEXTsymbol{\backslash}\TEXTsymbol{\backslash}%
ftp://casa.usno.navy.mil/navnet/postscript/prd\TEXTsymbol{\backslash}\_15.ps
(1999).

\bibitem[FoSr1976]{Fomalont 1976 VLBI gamma determination}  E.B. Fomalont and 
R.A. Sramek. \textit{Measurements of the Solar gravitational deflection of 
radio waves in agreement with General Relativity.} Physical Review Letter, 36, 
1475-1478 (1976).

\bibitem[GAIA2000]{GAIA 2000 study repport}  GAIA Science Advisory Group. 
\textit{GAIA: Composition, formation and evolution of the galaxy. Results of
the Concept and Technology Study}.
http://astro.estec.esa.nl/SA-general/Projects/GAIA/, march-report.pdf,
version 1.6 (2000).

\bibitem[GrNa1989]{Grossman 1989 Lensing arcs clusters}  S. A. Grossman and
R. Narayan. \textit{Gravitationally lensed images in ABBEL 370}.
Astrophysical Journal, 344, 637-644 (1989).

\bibitem[Le et al.1995]{Lebach 1995 VLBI}  D. E. Lebach, B. E. Corey, I. I.
Shapiro, M. I. Ratner, J. C. Webber, A. E. E. Rogers, J. L. Davis and T. A.
Herring. \textit{Measurements of the solar gravitational deflection of radio
waves using Very Long Baseline Interferometry}. Physical Review Letters, 75,
8, 1439-1442 (1995).

\bibitem[Na1997]{Narayan 1997 lectures gravitational lensing}  R. Narayan. 
\textit{Lectures on gravitational lensing}. Harvard-Smithsonian Center for
Astrophysics, astro-ph/9606001 v2 (1997).

\bibitem[Ma1994]{Mannheim 1994 microlensing}  P. D. Mannheim. \textit{%
Microlensing, Newton-Einstein gravity, and conformal gravity.} UCONN-94-10,
astro-ph/9412007 (1994).

\bibitem[Ma1995]{Mannheim 1995 Age of Universe}  P. D. Mannheim. \textit{%
Conformal Cosmology and the Age of the Universe}. UCONN 95-08 (1995).

\bibitem[MaKa1989]{Mannheim 1989 Exact Solution}  P. D. Mannheim and D.
Kazanas. \textit{Exact Vacuum Solution to Conformal Weyl Gravity and
Galactic Rotation Curves}. Astrophysical Journal, 342, 635-638 (1989).

\bibitem[Pe et al.1991]{Pello 1991 Lensing arcs clusters}  R. Pello and
J.-F. Le Borgne and G. Soucail and Y. Mellier. \textit{A straight
gravitational image in ABBEL 2390: a striking case of lensing by a cluster
of galaxies}. Astrophysical Journal, 366, 405-411 (1991).

\bibitem[Pi1997]{Pireaux 1997 memoire}  S. Pireaux. \textit{Etude de
Solutions Particuli\`{e}res d'une Th\'{e}orie Invariante Conforme de la
Gravitation}. Graduate thesis, Universit\'{e} catholique de Louvain (UCL),
June 1997.

\bibitem[Pi2002]{Pireaux 2002 these}  S. Pireaux. \textit{Light deflection
experiments as a test of relativistic theories of gravitation}. Ph.D.
thesis, Universit\'{e} catholique de Louvain (UCL), September 2002.

\bibitem[Pi2004]{Pireaux 2003 critical distances in Weyl}  S. Pireaux. 
\textit{Light deflection in Weyl gravity: critical distances for photon
paths.} Classical and Quantum Gravity, 21, 1897-1913 (2004), gr-qc/0403071.

\bibitem[RoCa1984]{Robertson 1984 VLBI gamma measurements}  D.S. Robertson 
and W.E. Carter.  \textit{Relativistic deflection of radio signals in the solar 
gravitational field measured with {VLBI}.} Nature,  310, 572-574 (1984).

\bibitem[Ro et al.1991a]{Robertson 1991 VLBI gamma measurements} D.S. Robertson, 
W.E. Carter and W.H. Dillinger. \textit{New measurements of solar 
gravitational deflection of radio signals using {VLBI}.} Nature, 349, 768-770 
(1991).

\bibitem[Ro et al.1991b]{Robertson 1991 VLBI gamma measurements proceedings} 
D.S. Robertson, W.E. Carter and W.H. Dillinger. \textit{A New measurement of solar 
gravitational deflection of radio signals using {VLBI}.} Proceedings of the A.G.U. 
Chapman Conference: "Geodetic {VLBI}: monitoring global change", Washington, April 1991, 
22-26, 203-212 (1991).

\bibitem[Se et al.1970]{Seielstad 1970 VLBI gamma determination}  G.A. Seielstad, 
R.A. Sramek and K. Weiler. \textit{Measurement of the deflection of {9.602-GHz} 
radiation from {3C279} in the solar gravitational field.} Physical Review Letter, 
24, 1373-1376 (1970).

\bibitem[Sh et al.2004]{Shapiro 2004 VLBI estimate of gamma}  S. S. Shapiro, 
J.L. Davis, D.E. Lebach and J.S. Gregory. \textit{Measurement of the solar 
gravitational deflection of radio waves using geodetic {Very-Long-Baseline 
Interferometry} data, 1979-1999.} Physical Review Letters, 92, 121101 (2004).

\bibitem[Turn1999]{Turner 1999 Estimations of Omega}  M. S. Turner. \textit{%
Dark matter, dark energy and fundamental physics}. Proceedings of ``Physics
in Collision'', Ann Arbor, MI, 24-26th June 1999, Eds. M. Campbell, T. M.
Wells (World Scientific, NJ). astro-ph/9912211.

\bibitem[Tury et al.2004]{Turyshev 2003 LATOR}  S. G. Turyshev, M. Shao and
K. Jr. Nordtvedt. \textit{The Laser Astrometric Test of Relativity (LATOR)
mission}. Astronomische Nachrichten, 325, 267-277 (2004). 
gr-qc/0311020 v1, gr-qc/0311049 v1.

\bibitem[Wi2001]{Will 2001 summary of tests}  C. M. Will. The confrontation
between General Relativity and experiments. Living Reviews, 4
(2001). \TEXTsymbol{\backslash}\TEXTsymbol{\backslash}%
www.livingreviews.org/Articles/Volume4/2001-4will
\end{thebibliography}
\end{document}